\documentclass[conference]{IEEEtran}

\IEEEoverridecommandlockouts
\IEEEpubid{\makebox[\columnwidth]{978-1-7281-8086-1/20/\$31.00 \copyright~2020 IEEE \hfill} \hspace{\columnsep}\makebox[\columnwidth]{ }}

\usepackage[binary-units=true]{siunitx}
\DeclareSIUnit \dBm {dBm}
\DeclareSIUnit \dB {dB} 
\DeclareSIUnit \dBi {dBi} 
\DeclareSIUnit \Kbps {Kbps}
\DeclareSIUnit \Mbps {Mbps}
\DeclareSIUnit \Gbps {Gbps}
\DeclareSIUnit \kBps {kBps}
\DeclareSIUnit \MBps {MBps}
\DeclareSIUnit \GBps {GBps}

\usepackage{cite}   
\usepackage{makecell}
\usepackage[caption=false, font=footnotesize]{subfig}
\usepackage[dvips]{graphicx}
\usepackage{url}       
\usepackage{amsfonts}
\usepackage{amssymb}
\usepackage{mathtools}
\usepackage{times}
\usepackage{algpseudocode}
\usepackage{fourier}
\usepackage{algorithm}
\usepackage{enumerate}
\usepackage{enumitem}
\usepackage{multirow}
\usepackage{tikz}
\usepackage{array}
\usepackage{gensymb}
\usepackage{eucal}
\usepackage{mathtools}
\usepackage{nicefrac}
\usepackage{tabularx}

\newcolumntype{P}[1]{>{\centering\arraybackslash}p{#1}}

\bstctlcite{IEEEexample:BSTcontrol}

\algnewcommand\algorithmicinput{\textbf{Level 1:}}
\algnewcommand\Level{\item[\algorithmicinput]}

\algnewcommand\algorithmicinputt{\textbf{Level 2:}}
\algnewcommand\Levell{\item[\algorithmicinputt]}

\algnewcommand\algorithmicinputtt{\textbf{Level 3:}}
\algnewcommand\Levelll{\item[\algorithmicinputtt]}

\algnewcommand\algorithmicinputttt{\textbf{Output:}}
\algnewcommand\Output{\item[\algorithmicinputttt]}

\usepackage[english]{babel}
\usepackage{blindtext}
\IEEEoverridecommandlockouts
\begin{document}
\title{DRIVE: A Digital Network Oracle for Cooperative Intelligent Transportation Systems}

\author{\IEEEauthorblockN{Ioannis Mavromatis\IEEEauthorrefmark{1}\IEEEauthorrefmark{2}, Robert J. Piechocki\IEEEauthorrefmark{2}\IEEEauthorrefmark{3}, Mahesh Sooriyabandara\IEEEauthorrefmark{1}, and Arjun Parekh\IEEEauthorrefmark{4}}
\IEEEauthorblockA{\IEEEauthorrefmark{1} Bristol Research and Innovation Laboratory (BRIL), Toshiba Europe Ltd., Bristol, UK\\
\IEEEauthorrefmark{2} Department of Electrical and Electronic Engineering, University of Bristol, Bristol, UK\\
\IEEEauthorrefmark{3}The Alan Turing Institute, London, UK\\
\IEEEauthorrefmark{4} British Telecom (BT) Group, Bristol, UK\\
Email: Ioannis.Mavromatis@toshiba-trel.com, R.J.Piechocki@bristol.ac.uk, Mahesh@toshiba-trel.com, Arjun.Parekh@bt.com}}

% \IEEEaftertitletext{\vspace{-3mm}}
\maketitle
\IEEEpubidadjcol

\begin{abstract}
In a world where Artificial Intelligence revolutionizes inference, prediction and decision-making tasks, Digital Twins emerge as game-changing tools. A case in point is the development and optimization of Cooperative Intelligent Transportation Systems (C-ITSs): a confluence of cyber-physical digital infrastructure and (semi)automated mobility. Herein we introduce Digital Twin for self-dRiving Intelligent VEhicles (DRIVE). The developed framework tackles shortcomings of traditional vehicular and network simulators. It provides a flexible, modular, and scalable implementation to ensure large-scale, city-wide experimentation with a moderate computational cost. The defining feature of our Digital Twin is a unique architecture allowing for submission of sequential queries, to which the Digital Twin provides instantaneous responses with the “state of the world”, and hence is an Oracle.  With such bidirectional interaction with external intelligent agents and realistic mobility traces, DRIVE provides the environment for development, training and optimization of Machine Learning based C-ITS solutions.  
%In a world where Artificial Intelligence revolutionizes decision making, Digital Twins can prove game-changing for the future of Cooperative Intelligent Transportation Systems (C-ITSs). Digital Twins can enhance experimental validations, representing accurately and in a matter of seconds, hundreds of vehicles and the interactions between them. In our work, we introduce Digital twin for self-dRiving Intelligent VEhicles (DRIVE). Our framework tackles existing problems of the traditional vehicular simulators. It provides a flexible, modular, and scalable implementation to ensure large-scale experimentation in minimal time. Providing a bidirectional interaction with intelligent agents and realistic mobility traces, DRIVE can simulate different C-ITS aspects and provides the foundation for several use-case demonstrations. With DRIVE, we envision to advance the state-of-the-art in performance evaluation of C-ITSs and enhance the prototyping of intelligent solutions. 
\end{abstract}

\vspace{-1mm}
\section{Introduction}

Next-generation Cooperative Intelligent Transportation Systems (C-ITSs) will bring the paradigm of Mobility-as-a-Service (MaaS) to a whole new level~\cite{maas}. The C-ITS infrastructure network and the Connected and Autonomous Vehicles (CAVs) will share sensor data and maneuvering intentions in a heterogeneous Vehicle-to-Everything (V2X) fashion~\cite{hetNet}. These interactions can enhance the environmental perception, enable co-operative driving and the whole host of additional mobility services~\cite{cavs}.

The connectivity performance within C-ITSs and CAVs has been extensively researched~\cite{cavsSurvey,itsSurvey}. The lack of real CAVs and C-ITSs for experimentation is tackled through complex simulation frameworks~\cite{parallelInet}. Such an approach can reduce the immense cost of using large fleets of vehicles, and provides an easy and repeatable way to obtain near-perfect results. Based on the above, the concept of ``Digital Twins'' has been proposed in the past~\cite{digitalTwins}. A Digital Twin attempts to replicate salient features and data flows of the real world for the problem at hand. 

Traditional wireless networking problems are recently tackled with Machine Learning (ML) techniques and advanced simulation models~\cite{mlAir}. Traditional ML usually relies on existing datasets to train and evaluate a model. Later, the trained model is deployed on the end-system. However, C-ITSs can evolve over time, thus requiring more sophisticated decision agents that interact with the environment in real-time (e.g., Reinforcement Learning (RL) agents). Based on that, in this paper, we introduce the concept of a ``Digital Network Oracle'', an augmented ``Digital Twin''. A ``Digital Network Oracle'' introduces a particular architecture that allows the Twin to be queried sequentially, returning a snapshot of the state of the ``virtual world''. With such capabilities, the Twin can be used to train RL controllers for a swath of inferential and prediction tasks. A Digital Network Oracle, moves beyond static data and assets, and accesses interactions between entire systems covering vehicles, people, processes, and behaviors. Based on the above concepts, we developed our own Digital Network Oracle, called \textit{Digital twin for self-dRiving Intelligent VEhicles (DRIVE)}, and we introduce it in the rest of the paper. Our framework is publicly available under {\tt\small github.com/ioannismavromatis/DRIVE\_Simulator}.

%Traditional simulations do not provide insights into the interactions with the physical world. Digital twins can provide improved designs linking digital models and simulations with real-world data.
%On the other hand, Digital Twins are suitable for playing out the \textit{what-if} scenario for an asset. In the real world, when assets (e.g., an autonomous vehicle) intersect with other assets, people, and processes, that arises many undesired and unexpected behaviors. 
% A Digital Network Oracle, being the evolution of Digital Twins, moves beyond static data and assets and accesses interactions between entire systems covering vehicles, people, processes, and behaviors. As our proposed Digital Network Oracle, we introduce the \textit{Digital twin for self-dRiving Intelligent VEhicles (DRIVE)}. 

C-ITSs, are a great example of a System-of-Systems (SoS), and are information-centric, i.e., transport and processing of data and information in integral to their performance and dependable operations. With that in mind, we see that traditional simulation frameworks lack in several areas. For example, we observe a lack of supporting technologies (e.g., Veins framework~\cite{veins} supports only IEEE 802.11p). Other frameworks significantly increase the computational complexity (e.g., Veins, iTetris~\cite{iTetris}, etc). Frameworks that address the complexity (e.g., Vienna 5G System Level Simulator~\cite{Vienna5GSLS}), lack in the bidirectional interactions with real-world mobility traces, or modularity and extensibility. 

%Traditional frameworks lack in several aspects. For example, Veins framework~\cite{veins} supports only IEEE 802.11p as a communication plane. What is more, Veins, INET, iTetris~\cite{iTetris}, and other similar frameworks significantly increase the computational complexity, and evaluating large-scale scenarios becomes a demanding task. Frameworks like Vienna 5G System Level Simulator~\cite{Vienna5GSLS} and Geometry-based Efficient propagation Model for V2V communication (GEMV$^2$) framework~\cite{gemv2} reduce the computational complexity with the cost of missing features (e.g., bidirectional interactions with real-world-like mobility traces). C-ITSs, like all System-of-Systems (SoS), are, in fact, information-centric, i.e., transport and processing of data and information are central to their performance and dependable operations. At the same time, they rely on several communication planes to extend their networked connectivity towards a wide device range. 

DRIVE is designed with the above in mind and can address several shortcomings of the existing frameworks. More specifically, we introduce a flexible and modular implementation, so the required level of realism is chosen, without wasting essential computational resources. Flexible evaluation mechanisms are developed that report systemic changes and behaviors after a set of interactions. What is more, during the execution of the scenarios, bidirectional interaction with realistic mobility traces is provided~\cite{sumo}. Finally, uncertainty modeling and the unpredictability found in the real-world scenarios are introduced within the framework. Overall, DRIVE, addressing the above, can be used for holistically tackling C-ITSs problems.

The rest of the paper is organized as follows. In Sec.~\ref{sec:requirements} we describe the key requirements for a Digital Network Oracle, its advantage, and potential use-cases. Sec.~\ref{sec:drive} describes the design and the implementation of the main DRIVE components and the benefits introduced. A reference scenario is described and evaluated in Sec.~\ref{sec:evaluation}. Finally, the paper is concluded in Sec.~\ref{sec:conclusions} with some suggestions for the future.

\section{Requirements for a Digital Network Oracle}\label{sec:requirements}
\vspace{-1mm}
Overall, a Digital Network Oracle provides a framework where intelligent agents can coexist with a realistic environment and interact in real time. Some potential use-cases when merging ML/RL and a Digital Network Oracle are:
\begin{itemize}
    \item \textbf{Anomaly detection}: A sequence of observations is examined to identify points in the time series where unusual events may have occurred.
    \item \textbf{Performance analysis}: The accumulated Quality-of-Service (QoS) is of prime concern for C-ITSs. The accumulation horizon can vary, and individual rewards can describe the behavior. A typical use case might be a systematic analysis of different wireless environments.
    \item \textbf{Stress test analysis}: Manufacturing of a non-stationary distribution, designed to place an increasing demand on the system. In such a case, one would attempt to observe a QoS metric for possible failure points.    
    % \item \textbf{RL agent training}: An RL agent interacts with the Twin and attempts to devise an optimal policy to maximize some reward (e.g. QoS metric).
\end{itemize}{}

The above use-cases can be tackled by ML/RL techniques~\cite{mlAir}. When combined, though, with the highly diverse vehicular environments, the required interactions introduce new obstacles in the existing simulation approaches. For a successful C-ITS evaluation, all the subsystems in this SoS (human users, vehicles, communication planes, etc.) should interact flawlessly and in real-time. Based on the above, some key requirements for a Digital Network Oracle are: 

\begin{enumerate}[wide, labelwidth=!, labelindent=9pt]
    \item \textbf{\textit{Repeatability}}: It must serve as a real-time environment, generating information for an intelligent agent, and applying new policies. Randomness can potentially be introduced for further validation and increased realism, however, it must be controlled by the user (e.g., with seeds).\label{req:req1}

    \item \textbf{\textit{Flexibility and Modularity}}: New systems or subsystems should be easily composed (e.g., a new communication plane could be introduced with a change in the configuration parameters). Different parameters must be exposed to adjust between the realism and the execution speed. For example, adjusting the map resolution could benefit the rapid prototyping and debugging (when decreased) or enhance the realism (when increased).\label{req:req2}
    
    \item \textbf{\textit{Bidirectional Interaction with Intelligent Agents}}: The framework should handle the training data as well as receive and apply the policies generated by intelligent agents. The framework should provide corresponding data interfaces for interaction with existing libraries.\label{req:req3}
    
    \item \textbf{\textit{Bidirectional Interaction with Traffic Management Tools and Maps}}: The framework must provide bidirectional interaction with mobility models (both vehicles and pedestrians) and manipulation of real-world maps. Common formats should be agreed between the traces and the maps.\label{req:req4}
    
    \item \textbf{\textit{Robust evaluation mechanisms}}: All intelligent agents should be linked with existing evaluation mechanisms. The status of assigned tasks (e.g., finished, pending, errors) should be exposed as well as the change in the performance introduced by a new policy. \label{req:req5}
    
    \item \textbf{\textit{Parallel sampling ability}}: The agent should query the environment (exploration) and adjust itself by the information and feedback received. Parallel sampling could be provided to increase the training efficiency. \label{req:req6}
    % To train an intelligent agent,the  exploration  of  the  environment  is  necessary.  The  agent should  query  the  environment  and  adjusting  itself  by  the information  and  feedback  received.  If  parallel  sampling  is possible for the simulation framework, the training process of  the  agent  would  be  more  efficient. \label{req:req6}
    
    \item \textbf{\textit{Minimal overhead and faster execution}}: 
    The framework must mitigate the computational complexity by introducing minimal overhead and ways of exploiting all the system resources (e.g., parallelization of tasks).\label{req:req7} 
    % Training intelligent agents usually dictate the computational resources and introduce bottlenecks, especially if being a single component. The framework should mitigate this problem by introducing minimal overhead and ways of exploiting all the system resources (e.g., parallelization of tasks).\label{req:req7}
    
\end{enumerate}

\begin{figure*}[t]     
\centering
    \includegraphics[width=1\textwidth]{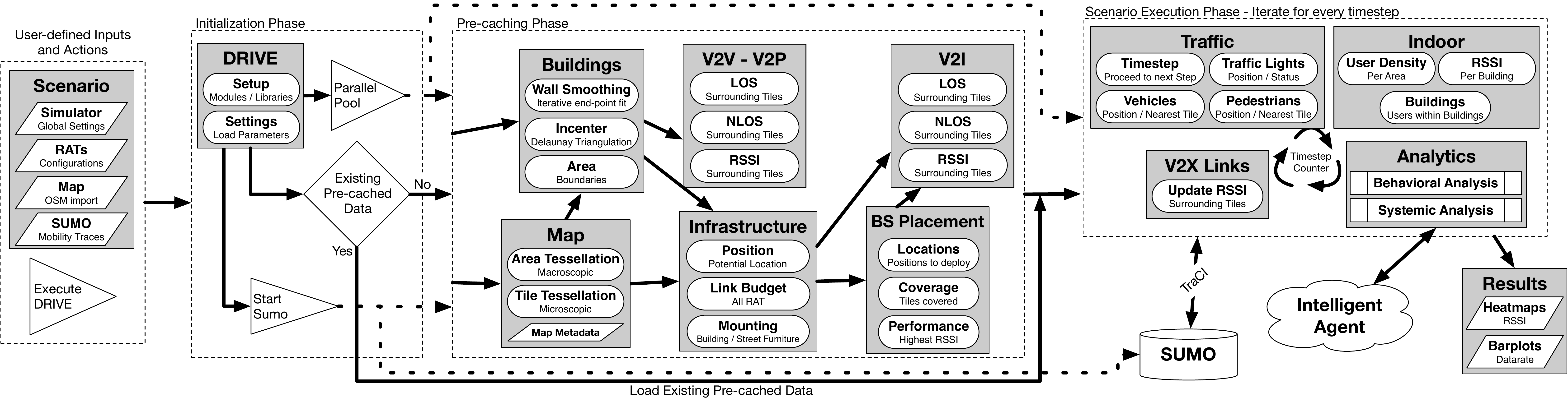}
    \caption{A high-level framework design of DRIVE, with the steps followed during the initialization, pre-caching, and scenario execution phases. The user inputs several parameters via configuration files, and the framework handles the different interactions later. A pre-caching phase can speed up the consecutive executions, while interactions with mobility traces and intelligent agents can introduce new avenues of realistic C-ITS evaluation.}
    \label{fig:high-level}
\end{figure*}

\vspace{-1mm}
\subsection{Limitations of Existing Vehicular Simulation Frameworks}\label{sub:existing}

There are several well-known simulation frameworks used for vehicular communications and C-ITS scenarios. The features of these frameworks vary in terms of the coupling (unidirectional or bidirectional) between mobility traces (realistic or artificially generated) and vehicular communication planes. With regards to the communication links, they are either described in a fine-grained fashion (packet-level) or focus more on the global behavior of large cyber-physical systems (system-level). Table~\ref{tab:simulators} gives an overview of the existing popular frameworks.

% Modern C-ITS scenarios require the integration and interoperability of several communication over one traffic management plane. Many of the well-known vehicular simulators provide a unidirectional or bidirectional coupling between realistic or artificially generated mobility traces and a communication framework. Each tool provides one or a number of supported communication technologies. These technologies are either described in a fine-grained fashion (packet-level) or focus more on the global behavior of large cyber-physical systems (system-level).  Tab.~\ref{tab:simulators} gives an overview of the existing popular frameworks.

\begin{table*}[t] 
\renewcommand{\arraystretch}{1.06}
\centering
    \caption{State-of-the-art Frameworks for Simulation of Vehicular Communications and Comparison with DRIVE.}
    \begin{tabular}{|r||c|c|c|c|c|c|c|c|}
    \hline \textbf{\makecell[r]{Simulation \\ Framework}} & \textbf{\makecell{Open \\ Source}} & \textbf{\makecell{Mobility \\ Interaction}} & \textbf{\makecell{Traces -- \\Maps}}  & \textbf{Level} & \textbf{\makecell{Supported Wireless \\ Technologies}} & \textbf{\makecell{Core\\ Framework}} & \textbf{\makecell{Intel. Agent\\ Interaction}} & \textbf{Scalability}   \\ \hline \hline
    Veins~\cite{veins} & Yes & Bidirectional  & Real-world & Packet & IEEE 802.11p & OMNeT++ & No  & Poor \\ \hline
    INET & Yes & Bidirectional & Real-world &  Packet & IEEE 802.11x, Cellular & OMNeT++ & No  & Poor \\ \hline
    iTetris~\cite{iTetris} & Yes & Bidirectional & Real-world  & Packet & IEEE802.11p, WiMAX & NS3 & No  & Poor \\ \hline
    Netsim~\cite{netsim} & No & Unidirectional & Artificial & Packet & IEEE 802.11p & Standalone & No  & Poor \\ \hline
    TraNS~\cite{trans} & Yes & Bidirectional & Real-world & Packet & Generic Wireless Plane & NS2 & No  & Poor  \\ \hline
    VSimRTI~\cite{vsimrti} & No & Bidirectional & Real-world & Packet & IEEE 802.11p, Cellular & OMNeT++, NS3 & No & Poor  \\ \hline
    Vienna 5G~\cite{Vienna5GSLS} & Yes & Unidirectional & Artificial & System & Cellular/5G & Standalone & No & Fair \\ \hline
    GEMV$^2$~\cite{gemv2} & Yes & Unidirectional & Real-world & System & IEEE 802.11p & Standalone & No & Good  \\ \hline
    DRIVE & Yes & Bidirectional & Real-world & System & Technology-agnostic & Standalone & Yes & Great \\\hline
    \end{tabular}
\label{tab:simulators}
\end{table*}

Starting with the packet-level network simulators, and as shown in Table~\ref{tab:simulators}, all frameworks except Netsim~\cite{netsim}, provide integration with SUMO traffic generator~\cite{sumo}, whereas, Netsim artificially generates the mobility traces used.
% Well-known, open-source, packet-level network simulators are the INET/Veins~\cite{veins}, iTetris~\cite{iTetris}, and TraNS~\cite{trans}, based on OMNeT++~\cite{omnetpp}, ns-3~\cite{ns3} and ns-2 network simulators. Similar closed-source software suites are Netsim~\cite{netsim} and VSimRTI~\cite{vsimrti}. All the frameworks, except Netsim, provide integration with SUMO traffic generator~\cite{sumo}. Netsim mobility traces are artificially generated and do not reflect the reality. 
Furthermore, Veins~\cite{veins}, iTetris~\cite{iTetris} and VSimRTI~\cite{vsimrti} allow online re-configuration and re-routing of vehicles in reaction to network packets. Finally, all packet-level simulators implement all or a number of the Open Systems Interconnection (OSI) layers.  That allows for fine-grained experimentation and results in excellent agreement with the real world. However, as shown in~\cite{parallelInet}, due to the assessment of the propagation characteristics on a per-packet-basis, city-scale scenarios and just a few minutes of real-world time, could easily result in days of simulation time. 

The computational complexity is being addressed by Vienna~\cite{Vienna5GSLS} and GEMV$^2$~\cite{gemv2} frameworks. Both reduce the complexity introducing a simplified propagation assessment scheme. The Vienna simulator implements a concise  ``map'' feature. whereas, GEMV$^2$, can parse real-world maps and traffic traces, but lacks the bidirectional interaction with them, a feature found in Veins or iTetris for example. 

As a general observation, and based on the requirements introduced before, it is evident that the existing frameworks lack essential features required for a systemic C-ITS experimentation. For example, their functionality is tied to specific communication technologies (e.g., GEMV$^2$ is based on just IEEE 802.11p), or the integration of new communication planes becomes convoluted (e.g., integration of LTE within Veins). Additionally, several frameworks do not provide interactions with real-world maps and mobility traces (e.g., Vienna Simulator and Netsim), limiting the synergy of mobility and communication links. All frameworks but Veins, limit the mobility traces to just one type of vehicle and do not consider pedestrians or indoor users. Finally, many introduce increased complexity, thus running scenarios in conjunction with intelligent agents becomes a very time-consuming activity.

\section{Design and Implementation of DRIVE framework}\label{sec:drive}

The above limitations show the obstacles introduced when evaluating realistic and large-scale C-ITS scenarios with the existing simulation frameworks. This is the gap that DRIVE intents to fill. DRIVE is a framework where all decision processes, routes, and interactions are determined at runtime. It can enhance the simulation of large-scale C-ITS city scenarios, where all traffic participants, and the communication infrastructure interact through various means of communication links. The framework is written in MATLAB and is open-sourced to enable the design of extensions by other users. In addition, MATLAB provides integration with all common programming languages (Python, C++, etc.), so libraries written in them can be easily utilized within our framework. A high-level visualization diagram of the developed framework can be found in Fig.~\ref{fig:high-level}.

% This is the gap that DRIVE intents to fill, introducing new strategies that allow the evaluation the systemic behaviors for large-scale C-ITSs in a SoS fashion. For example, coupling the mobility behavior with the network simulator itself instead of specific communication technologies could benefit the C-ITS experimentation. This is especially important when different types of vehicles (e.g., regular and emergency vehicles), their interactions with a human user, and heterogeneous communication technologies should be evaluated together in a single simulation scenario. 

DRIVE introduces a pre-caching feature not found in other simulation frameworks that can significantly enhance the user experience (Fig.~\ref{fig:high-level}). More specifically, static information such as building positions, potential BS placement, and potential communication interactions are calculated on the first run and stored for future usage. This ensures that when a user re-runs a scenario, the execution time will be minimized. In the next sections, we describe the information that is pre-cached or calculated on every run and how DRIVE operates under different scenarios.

DRIVE operates in two modes. The first one does not consider the mobility of vehicles and pedestrians, and the scenarios are based on just a given OSM map. This mode is ideal for evaluating different communication solutions with simplistic user densities per city block, as in our work~\cite{basestationPlacement}, where a city-scale infrastructure placement scheme was proposed.
% This mode can be used to evaluate different communication solutions based on more simplistic user densities per city block. As an example, a city-scale infrastructure placement scheme was proposed before~\cite{basestationPlacement}, which reduces the number of deployed basestations (BSs) based on a priori specified QoS constraints.
The second mode provides a more realistic C-ITS implementation. It introduces a bidirectional interaction with SUMO traffic generator~\cite{sumo} for both vehicle and pedestrian traffic. SUMO  interacts with our framework via the TraCI4Matlab framework~\cite{traci4matlab}. As before, different agents can be designed that will interact with the environment in various ways and devise optimal policies (e.g., a cell switch-on switch-off mechanism can be designed based on the traffic density on the roads). In the next sections, we describe in more detail the different components of DRIVE.

\subsection{Building Manipulation}\label{sub:buildings}

The Line-of-Sight (LOS) and Non-LOS (NLOS) communication links are dictated by the environmental obstacles, these usually being the buildings within cities. The propagated signal severely attenuates when intersecting with an obstacle. Especially for higher frequencies (e.g., Millimeter Wave (mmWave) frequencies of $>\SI{28}{\giga\hertz}$), usually, a single intersection with a wall almost completely blocks the signal. Usually, the calculation of the signal attenuation requires increased computational resources and introduces bottlenecks for the existing simulation frameworks~\cite{parallelInet}. Within DRIVE we introduce a pre-caching phase (Fig.~\ref{fig:high-level}) and several building manipulation strategies (Fig.~\ref{fig:high-level}-\textit{Buildings}) to minimize the execution time, without the loss of accuracy.

\begin{figure}[t]     
\centering
\includegraphics[width=1\columnwidth]{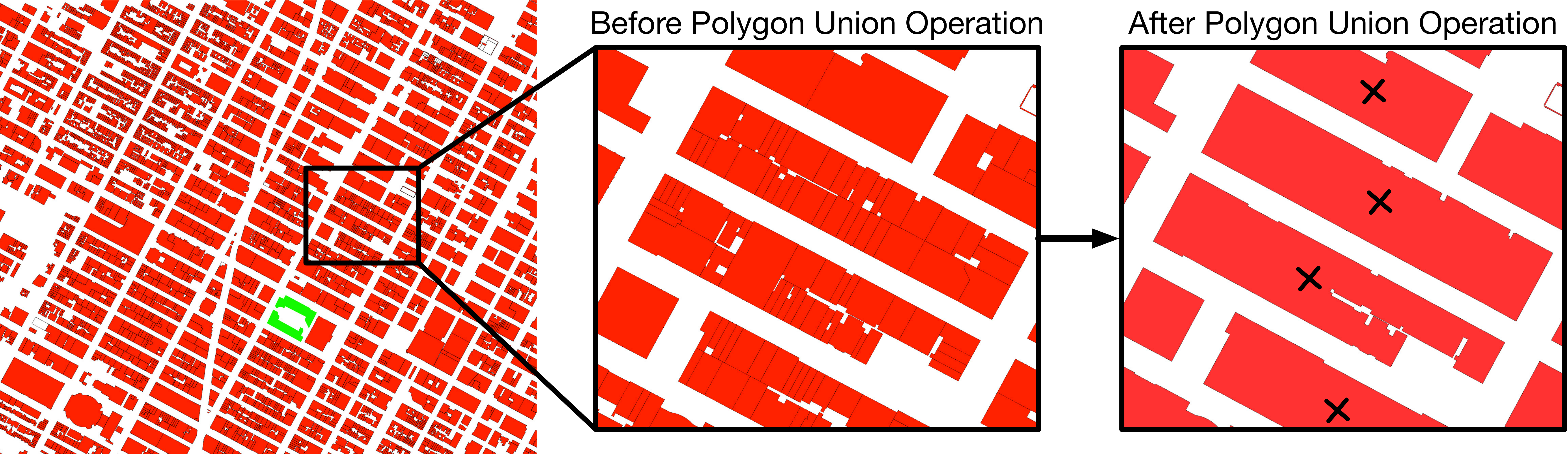}
    \caption{Example of a parsed OpenStreetMap file, the polygon union operation on two example city blocks and their incenters.}
    \label{fig:mapZoom}
\end{figure}

Under urban scenarios, building blocks consist of buildings with adjacent tangent sides or small negligible gaps between them (Fig.~\ref{fig:mapZoom}). Therefore, buildings on a map are represented as 2D or 3D \emph{Simple Polygons (SPs)}. An SP is considered a flat-shaped object consisting of straight, non-intersecting line segments. These lines, when joined pair-wise, form a closed path. The overall signal attenuation of a ray is the summation of the attenuation introduced by all intersecting building walls. Reducing the number of walls, the number of intersections is reduced as well. For DRIVE, we concatenate the side-by-side buildings using the polygon union operation~\cite{polygonUnion} and remove the holes introduced (e.g., a courtyard) to form solid objects (Fig.~\ref{fig:mapZoom}).

To further reduce the complexity, we introduce a smoothing of the buildings layout. Our approach decimates a curve composed of line segments, to a similar curve with fewer points based on a simplification tolerance parameter. Modifying this parameter, a user can decide the level of simplification required. For our implementation, we utilized the Douglas–Peucker iterative end-point fit algorithm. Reducing the number of polygon edges reduces the LOS and NLOS calculations for the entire city map. What is more, DRIVE introduces a 3D city environment. All building polygons inherit their height from OpenStreetMap (OSM) metadata~\cite{OpenStreetMap}. If the building height is missing, a value is assigned at random from a range controlled by the end-user. In our framework, there is also a provision to control the randomness with seeds to ensure the reproducibility of the results. The above simplifications align with Reqs.~\ref{req:req1}, ~\ref{req:req2} and~\ref{req:req7} introduced before. Based on the above, DRIVE can still achieve realistic performance from the system-level perspective, minimizing the computation time.

% The computational complexity is further reduced using the iterative end-point fit algorithm, or as otherwise known, the Douglas–Peucker algorithm. This approach decimates a curve composed of line segments, to a similar curve with fewer points based on a simplification tolerance parameter. Modifying this parameter, a user can decide the level of simplification required. Smoothing the polygon edges, the excessive corners are being removed, thus reducing the time needed to calculate the LOS and NLOS links for the entire city. OSM provides the building height for a number of large cities around the world. So, all the final building polygons inherit the height from the initial shape. When the height metadata are missing from OSM map, the user can input a range of values for all buildings to be assigned a height using uniform random distribution. Of course, the above feature is controlled by seeds to ensure the reproducibility of the results The above simplifications align with Reqs.~\ref{req:req1}, ~\ref{req:req2} and~\ref{req:req7} introduced before. Introducing them in DRIVE, we can still achieve a realistic performance investigation from the system-level perspective, minimizing the computational complexity of the excessive number of polygons. 

% \begin{figure}[t]     
% \centering
% \includegraphics[width=1\columnwidth]{unionPolygon.eps}
%     \vspace{-4mm}
%     \caption{Example of the polygon union operation for a given city block.}
%     \label{fig:unionPolygonOper}
% \end{figure}

\subsection{Interaction between Maps and Intelligent Agents}\label{sub:maps}
\vspace{-1mm}
As discussed in Sec.~\ref{sec:requirements}, one of the key requirements for a Digital Network Oracle is the bidirectional interaction with Intelligent Agents. Agents are responsible for generating policies that will be deployed on given scenarios. Lately, ``self-organized'' Multi-Agent Systems (MASs) are widely investigated. MAS is a network of agents who generate localized policies, having though as a goal to maximize the global reward.

\begin{figure}[t]     
\centering
\includegraphics[width=1\columnwidth]{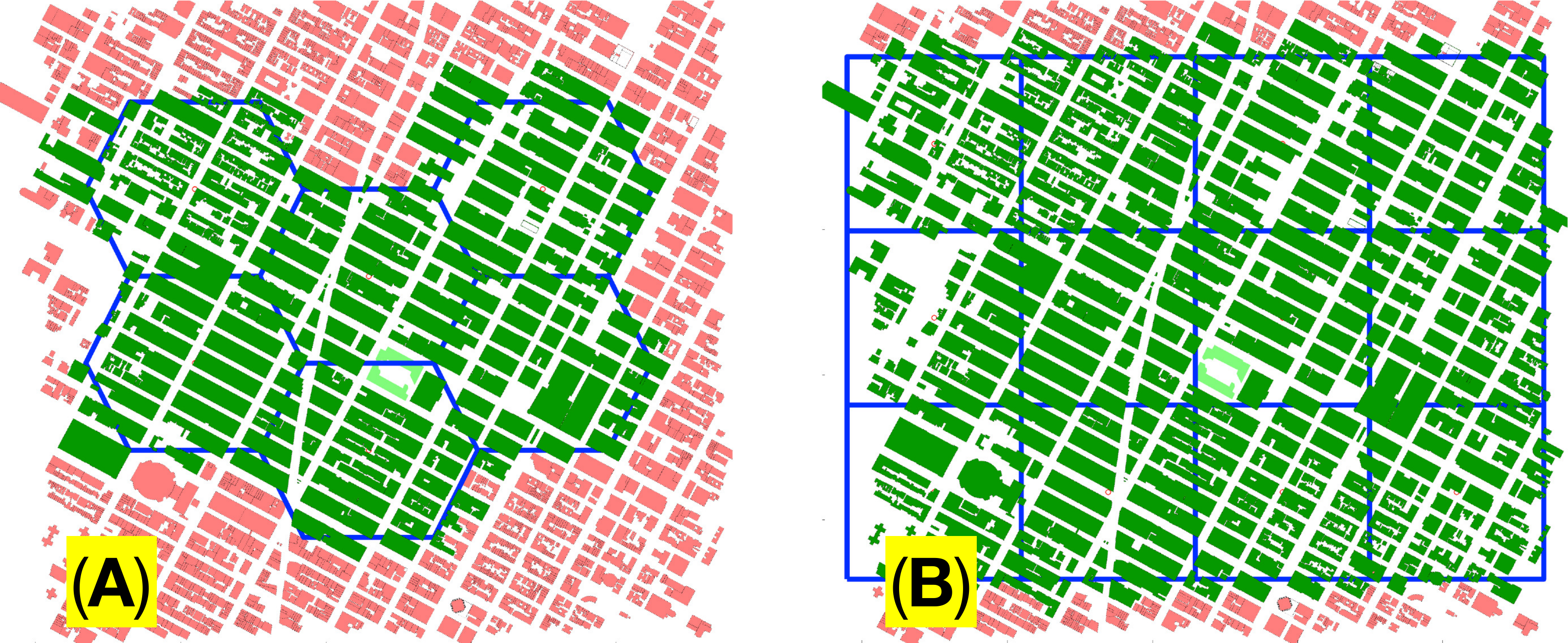}
    \caption{Examples of the tessellation approach. (A) shows the hexagonal tessellation while (B) shows the squared approach.}
    \label{fig:tessellation}
\end{figure}

A MAS-compliant architecture is established in DRIVE. We introduce a map tessellation in areas of equal size, in a two-way fashion (Fig.~\ref{fig:high-level}-\textit{Map}). Individual policies can be applied on each area. The areas can have either a squared or a hexagonal shape (Fig.~\ref{fig:tessellation}). The first is optimized for speed, while the latter provides a more realistic representation of the real world (e.g., the LTE BS coverage region is usually represented as a hexagonal pattern in the literature). All buildings with at least one edge within an area are considered part of this area. The benefits of such an approach are two-fold. Initially, all the interactions-of-interest can be focused on specific areas. That can reduce the execution time, without the need for reconfiguring the entire scenario. Most importantly, having a MAS intelligent scheme, we can apply different policies in different areas. Each individual agent can run locally and make decisions for specific areas. The above align with the Reqs.~\ref{req:req2},~\ref{req:req3},~\ref{req:req6} and~\ref{req:req7}. An example of that can be found in our previous work~\cite{basestationPlacement}.

\begin{figure}[t]     
\centering
\includegraphics[width=1\columnwidth]{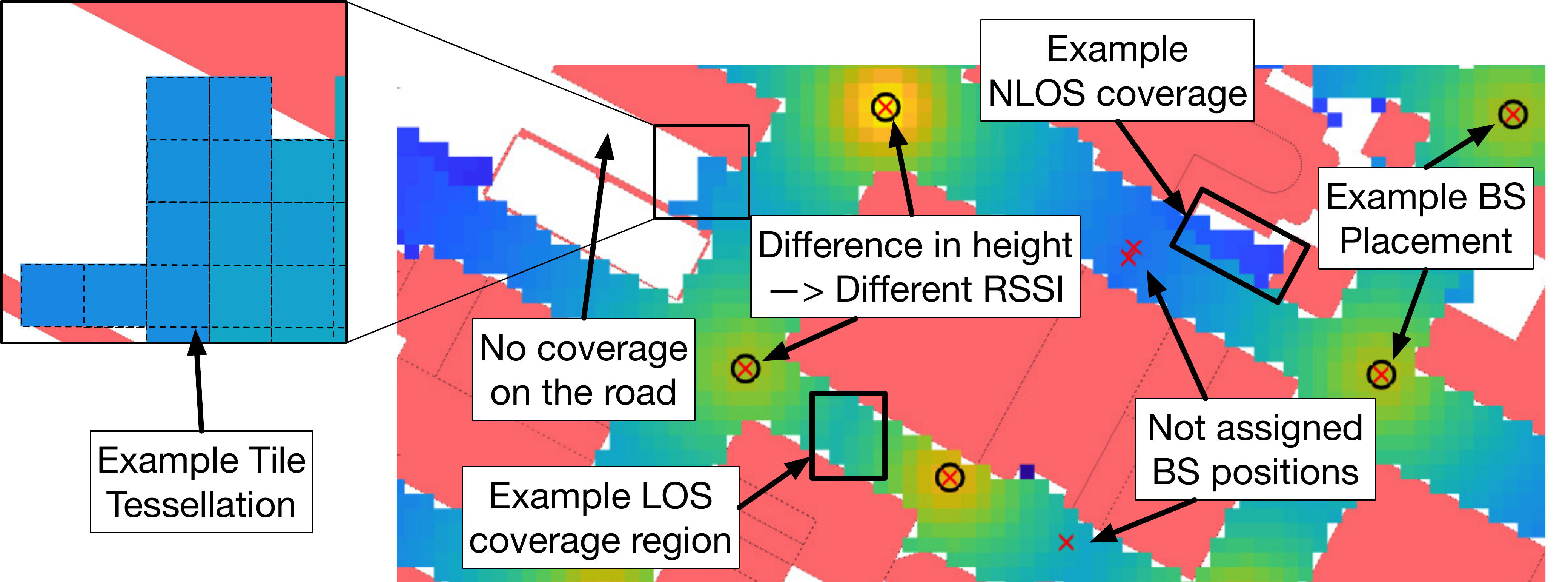}
    \caption{Example operational output. The heatmap shows the calculated RSSI per tile, the chosen femtocell BSs and the unassigned positions.}
    \label{fig:heatmap}
\end{figure}

A similar tessellation is followed in the microscopic world as well, with the map being discretized in tiles (Figs.~\ref{fig:heatmap} and~\ref{fig:high-level}-\textit{Map}). A tile is considered a small uniform area with the same properties on its entire surface. For DRIVE, all the calculations take place using the incenters of the given tiles to speed up the performance. Using a relatively small tile size (e.g., $\leq\SI{4}{\meter}$), a very realistic result can be achieved, significantly reducing the computational complexity (Req.~\ref{req:req7}). More information about that can be found in~\cite{basestationPlacement}.

\subsection{Communication Planes and Potential BS Positions}\label{sub:comms}
\vspace{-1mm}
DRIVE supports two ``cell-like'' types of communication planes. i.e., macrocell and femtocell, with the main difference being the mounting point of the BSs (Fig.~\ref{fig:high-level}-\textit{Infrastructure}). For the macrocell plane, the BSs are positioned on top of the buildings, placed at the incenter of the concatenated building blocks (Fig.~\ref{fig:mapZoom}). All buildings are considered potential BS positions during a simulation scenario (Fig.~\ref{fig:high-level}-\textit{BS Placement}). The macrocell plane is intended to be used for long-range high-power communication technologies (e.g., LTE, etc.).

\begin{figure}[t]     
\centering
    \includegraphics[width=1\columnwidth]{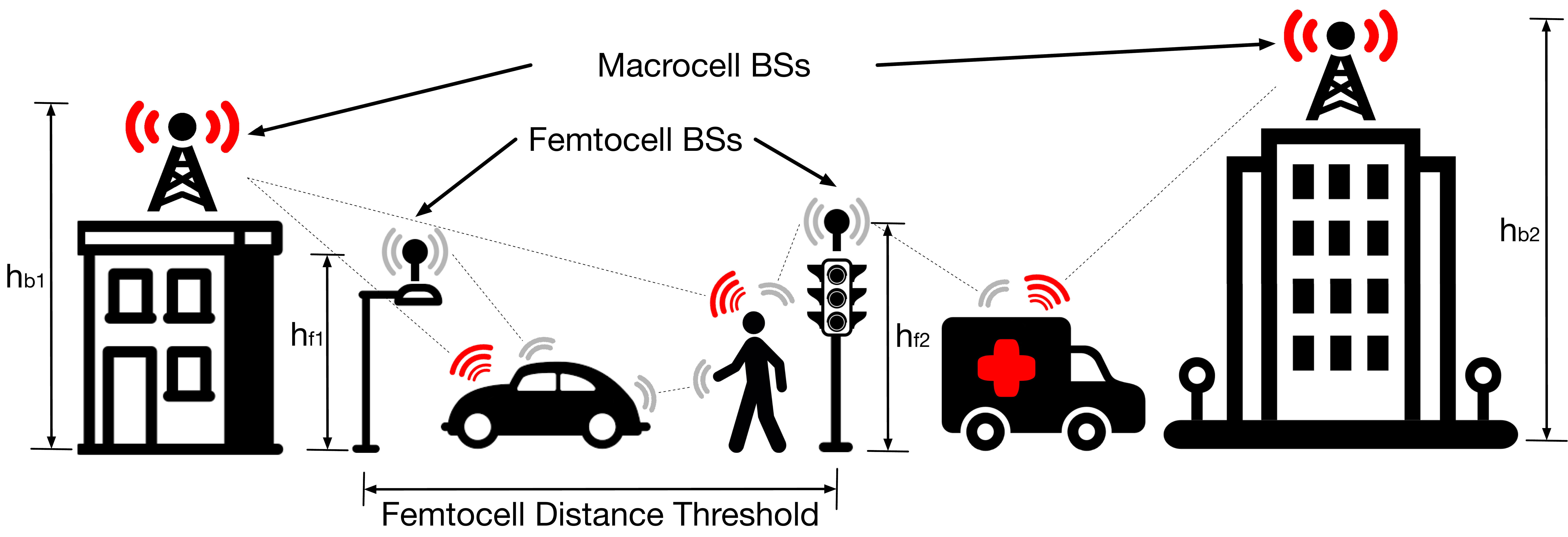}
    \caption{High-level system representation. The different types of vehicles and the pedestrians, equipped with a number of communication devices, choose the links that will be formulated based on an intelligent decisions.}
    \label{fig:system}
\end{figure}

The femtocell BSs are mounted on street furniture (Figs.~\ref{fig:heatmap} and~\ref{fig:system}). The traffic light locations are imported from the OSM map metadata. However, OSM does not provide information about lampposts positions (Fig.~\ref{fig:heatmap}). As a solution, for a given road, we artificially generate the lamppost positions by finding the road junctions and discretizing the road to equal segments, based on a user-defined value. The segments intersection point created are the potential lamppost positions (on both sides of the road). The above two sets of locations form the potential femtocell BS positions (Figs.~\ref{fig:heatmap} and~\ref{fig:high-level}-\textit{BS Placement})). The femtocell plane is intended to be used for short-range communication technologies (e.g., IEEE 802.11p, mmWaves, etc.).

The end-user can create different communication planes providing a configuration for each one with various wireless communication parameters for these cell types, i.e., carrier frequency, antenna gains, noise floor and figure, channel bandwidth, etc. DRIVE already implements some well known pathloss models (freespace, shadow fading described in~\cite{shadowFading}, and the urban macrocell model described in 3GPP TR 36.873) but users can define their own models as well (Fig.~\ref{fig:high-level}-\textit{Infrastructure}). The above provide a robust evaluation framework as the different performance indicators can be thoroughly evaluated.

Similarly, as in Sec.~\ref{sub:buildings}, the traffic lights, and the lampposts are assigned a height from a uniform random distribution within a given range of values (Fig.~\ref{fig:system}). At the moment, the vehicles are not considered as blocking objects. In the future, the 3D environment will be extended, with the blockage of moving vehicles as their height could affect the LOS or NLOS V2I links (Fig.~\ref{fig:high-level}-\textit{V2I}). All the above align with the Reqs.~\ref{req:req1}, ~\ref{req:req2}, and \ref{req:req5} mentioned before.

DRIVE provides a unique feature compared to many other frameworks. The objects, systems and interactions described in Secs.~\ref{sub:buildings},~\ref{sub:maps} and~\ref{sub:comms} are pre-cached before the interaction with a mobility model. Information such as the potential V2X interactions and their LOS/NLOS nature are pre-calculated using the existing information (Fig.~\ref{fig:high-level}-\textit{V2I/V2V-V2P}). Pre-calculating these information, and recalling them during the interaction with mobility traces can significantly reduces the execution time. To enhance the repeatability and the user experience, tools that store and load these information are provided. Utilizing such an approach, the scenario can run significantly faster after the first time as all the required information are already pre-calculated. These features align with the Reqs.~\ref{req:req1} and ~\ref{req:req2}.

% For example, coupling the mobility behavior with the network simulator itself instead of specific communication technologies could benefit the C-ITS experimentation. This is especially important when different types of vehicles (e.g., regular and emergency vehicles), their interactions with a human user, and heterogeneous communication technologies should be evaluated together in a single simulation scenario. 

\subsection{Interaction with Mobility Traces and Indoor Users}\label{sub:indoorOutdoor}

DRIVE can handle two sets of ``end-users''. Initially, DRIVE approximates the traffic density as a spatial Weibull distribution. The calculated distribution can be updated periodically generating realistic spatial traffic patterns for cellular networks can be achieved~\cite{indoorUsers}. The use of that is two-fold. When the vehicles and pedestrians are not considered, the generated distributions represent both indoor and outdoor users. On the other hand, when mobility traces are generated via SUMO, only the indoor user are based on this distribution.

% Initially we have the users per city block. DRIVE, based on the user input parameters, can demonstrate the traffic density (the traffic load per unit area) as a spatial distribution approximated by the Weibull distribution. The calculated distribution can be updated periodically generating realistic spatial traffic patterns for cellular networks can be achieved~\cite{indoorUsers}. This approach is used in two ways. When the vehicles and pedestrians are not considered, the generated distributions are used for the entire map (both indoor and outdoor connectivity). On the other hand, when the interaction with mobility traces is activated, only the indoor user are calculated based on this distribution.

For the latter, DRIVE provides a bidirectional interaction with the SUMO framework via TraCI4Matlab. The fuzziness in the SUMO route generation is controlled using seeds. During the execution time (Fig.~\ref{fig:high-level}-\textit{Traffic}), the position of all vehicles is acquired from SUMO, and the nearest tile incenter is calculated. Assigning vehicles to tile incenters, we can leverage from the pre-cached information for the V2X links and almost instantaneously find their expected V2X link performance. When updates happen in the network configuration (e.g., the TX power of a BS is increased), DRIVE updates the required potential links in real-time (Fig.~\ref{fig:high-level}-\textit{V2X Links}). Using the above architecture, DRIVE provides a flexible framework for interacting with intelligent agents and apply policies based on systemic behaviors. An example use-case can be the redirection of an emergency CAV from a particular route providing high-quality 5G connectivity to enable tactile robotic telesurgery. The above mentioned features align with the Reqs.~\ref{req:req1}, and \ref{req:req4}.

% This position is later correlated with the nearest tile incenter for each vehicle. Using the pre-cached information mentioned in Sec.~\ref{sub:comms}, the execution of this bidirectional interaction is almost instantaneous even for large number of vehicles (e.g., $>1000$ per timeslot). Based on all the above, intelligent agents can be designed that will gather information either for the systemic or the per-area traffic demand and manipulate the mobility based on intelligent decisions. An example use-case can be the redirection of a emergency CAV from a particular route providing high-quality 5G connectivity to enable a tactile robotic telesurgery. The above mentioned features align with the Reqs.~\ref{req:req1}, and \ref{req:req4}.

\section{Proof of Concept Evaluation}\label{sec:evaluation}
\vspace{-1mm}

\begin{table}[t]
\renewcommand{\arraystretch}{1.06}
\centering
\caption{Example Map Area and List of Simulation Parameters.}
\begin{tabularx}{.73\columnwidth}{*{1}{p{.20\columnwidth}}*{1}{P{.45\columnwidth}}}
\raggedleft\textbf{Urban Area} & \textbf{Manhattan, NY, USA} \\ \hline \hline
\raggedleft Center & $-73.9841\degree W, \, 40.75545\degree N$   \\
\raggedleft Map Size &  $\SI{2.274}{\kilo\meter} \times \SI{2.607}{\kilo\meter}$ \\
% \raggedleft Grid-Tile Size & $\SI{4}{\meter} \times \SI{4}{\meter}$  \\
\raggedleft Indoor Users & $\leq 100$ per building  \\
\raggedleft Outdoor Users & $200$ vehicles / $200$ pedestrians  \\\hline
\end{tabularx}
\newline
\vspace*{0.2cm}
\newline
\begin{tabularx}{\columnwidth}{*{1}{p{.36\columnwidth}}*{1}{P{.28\columnwidth}}*{1}{P{.26\columnwidth}}}
\raggedleft\textbf{Communication Plane} & \textbf{Macrocell}  & \textbf{Femtocell}  \\ \hline \hline
\raggedleft Example Technology   & LTE  & MmWaves  \\
\raggedleft Transmission power  & \SIrange{20}{43}{\dBm}  & \SIrange{15}{25}{\dBm}  \\
\raggedleft TX\,/\,RX Antenna Gain     & \SI{18}{\dBi}\,/\,\SI{0}{\dBi}  & \SI{22.6}{\dBi}\,/\,\SI{22.6}{\dBi} \\
\raggedleft Carrier Frequency   & \SI{2.6}{\giga\hertz}     & \SI{60}{\giga\hertz}  \\ 
\raggedleft Bandwidth 		    & \SI{20}{\mega\hertz}      & \SI{2.16}{\giga\hertz} \\ 
\raggedleft Propag. Model LOS\,/\,NLOS   & 3GPP TR 36.873\,/\, COST Hata Model  & Shadow Fading  \\
\raggedleft Path-Loss Exp. LOS\,/\,NLOS  & --      & 2.66\,/\,7.17    \\ 
\raggedleft Antenna Beamwidth  & $120\degree$      & $15\degree$    \\ 
% \raggedleft Noise Figure 		& $N_{\mathrm{fig}}$ & \SI{6}{\dB}  \\
% \raggedleft Noise Floor 		& $N_{\mathrm{fl}}$ & \SI{-174}{\dBm}  \\
% \raggedleft LOS Path Loss 		& $L_{\mathrm{LOS}}$ & $10 \, n \, \log_{10}(d) + C_{\mathrm{att}}$~\SI{}{\dB}  \\
\raggedleft Distance Separation  & -- & \SI{100}{\meter} \\
\raggedleft BS Height  & \SIrange{15}{50}{\meter} & \SIrange{5}{15}{\meter} \\ \hline
% \raggedleft Noise Power 		& $P_{\mathrm{n}}$ & $N_{\mathrm{fl}}+10\log_{10}B+N_{\mathrm{fig}}$~\SI{}{\dBm}  \\\hline
\end{tabularx}
\label{tab:simParameters}
\end{table}

In this section, we present the setup and the results of our proof of concept evaluation. Our reference map is a city area of Manhattan, USA. During the run time, the indoor users are picked from a random distribution, as discussed in Sec.~\ref{sub:indoorOutdoor}, while the outdoor users (vehicles and pedestrians) are generated using SUMO. We utilized two different communication planes, i.e., a macrocell plane based on 3GPP LTE, and a femtocell plane based on mmWaves. All the simulation parameters (chosen map, mobility traces, communication links) are summarized in Tab.~\ref{tab:simParameters}. As a scenario, we consider a simplistic intelligent agent that, based on the user density per timestep, adapts the TX power of the deployed BSs to enhance the datarate. 

\subsection{Evaluation of the Pre-caching Execution Time}
\begin{figure}[t]     
\centering
    \includegraphics[width=1\columnwidth]{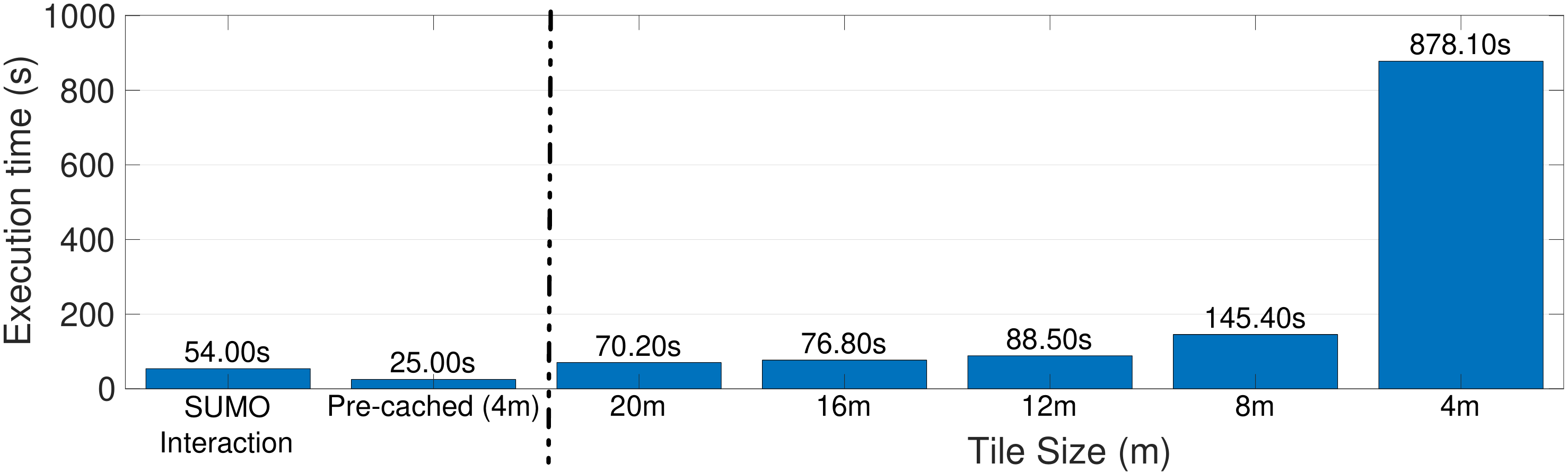}
    \caption{Execution time for different pre-caching scenarios. The first barplot shows the time required for the SUMO-DRIVE interaction (no other functions executed), the second shows the time needed for loading the pre-cached scenario information, and the rest show the overall time required for  the ``pre-caching phase'' and for different tile sizes.}
    \label{fig:times}
\end{figure}

\vspace{-1mm}
As discussed in Sec.~\ref{sub:existing}, one of the biggest drawbacks of the existing frameworks is their execution time. DRIVE, being highly optimized, manages to achieve rapid simulation time without loss in realism. In Fig.~\ref{fig:times}, we present the time required for pre-caching all the V2X LOS and NLOS interactions for our example scenario. We evaluated our framework for five different grid-like tessellation sizes, i.e., $\left\lbrace \SI{4}{}, \SI{8}{}, \SI{12}{}, \SI{16}{}, \SI{20}{} \right\rbrace\SI{}{\meter}$.  The simulated time is \SI{200}{\second}, and the mobility traces consist of 200 vehicles and 200 pedestrians controlled by SUMO. The chosen SUMO timestep length was \SI{1}{\second}. For our evaluation, we used a system with macOS 10.15, an 8-core Intel i9 \SI{2.3}{\giga\hertz} CPU, and \SI{16}{\giga\byte} of RAM. 

As expected, when the tessellation size increases, the pre-caching time increases as well. For fine-grained simulations (tile size $\leq\SI{4}{\meter}$), a big increase in the execution time is observed. This is due to the small RAM size of the used machine. For large-scale scenarios, like this one, and small tile sizes, at least \SI{32}{\giga\byte} of RAM is recommended. Overall, the execution time is always kept at reasonable levels. As shown, after the initial execution, the simulation length is significantly decreased. For example, the \SI{4}{\meter}-tile requires \SI{25}{\second} compared to the initial  \SI{878.1}{\second}. The pre-caching feature introduced can enhance the repeatability of the experimentation (e.g., when a bug is found in an under-development intelligent agent). The time needed during the scenario execution is significantly decreased as well. As shown \SI{200}{\second} of real-world time required roughly \SI{54}{\second} of simulation time within DRIVE. These results, compared to the hours of simulation time required by other frameworks~\cite{parallelInet}, can significantly enhance the end-user experience and reduce the overhead introduced.

\subsection{Example Use-case Scenario}\label{sub:scenario}
\vspace{-1mm}
In this section, we describe the actual execution of the scenario and how DRIVE handles the different interactions. We choose the BS on the map as follows (Fig.~\ref{fig:heatmap}). The macrocell BSs are equally spread with \textasciitilde\SI{500}{\meter} distance separation. Their exact location is the incenter of the nearest building block (Fig.~\ref{fig:mapZoom}). The femtocell BSs are placed using a greedy addition algorithm. Starting from the BS that covers the most tiles, we add BSs until $90\%$ of the city tiles are in LOS or NLOS coverage (as in~\cite{basestationPlacement}). 

\begin{figure*}[t]     
\centering
\begin{minipage}{.37\linewidth}
\centering
\subfloat[Snapshot of vehicles and pedestrians positions.]{\label{fig:mobility}\includegraphics[width=0.98\linewidth]{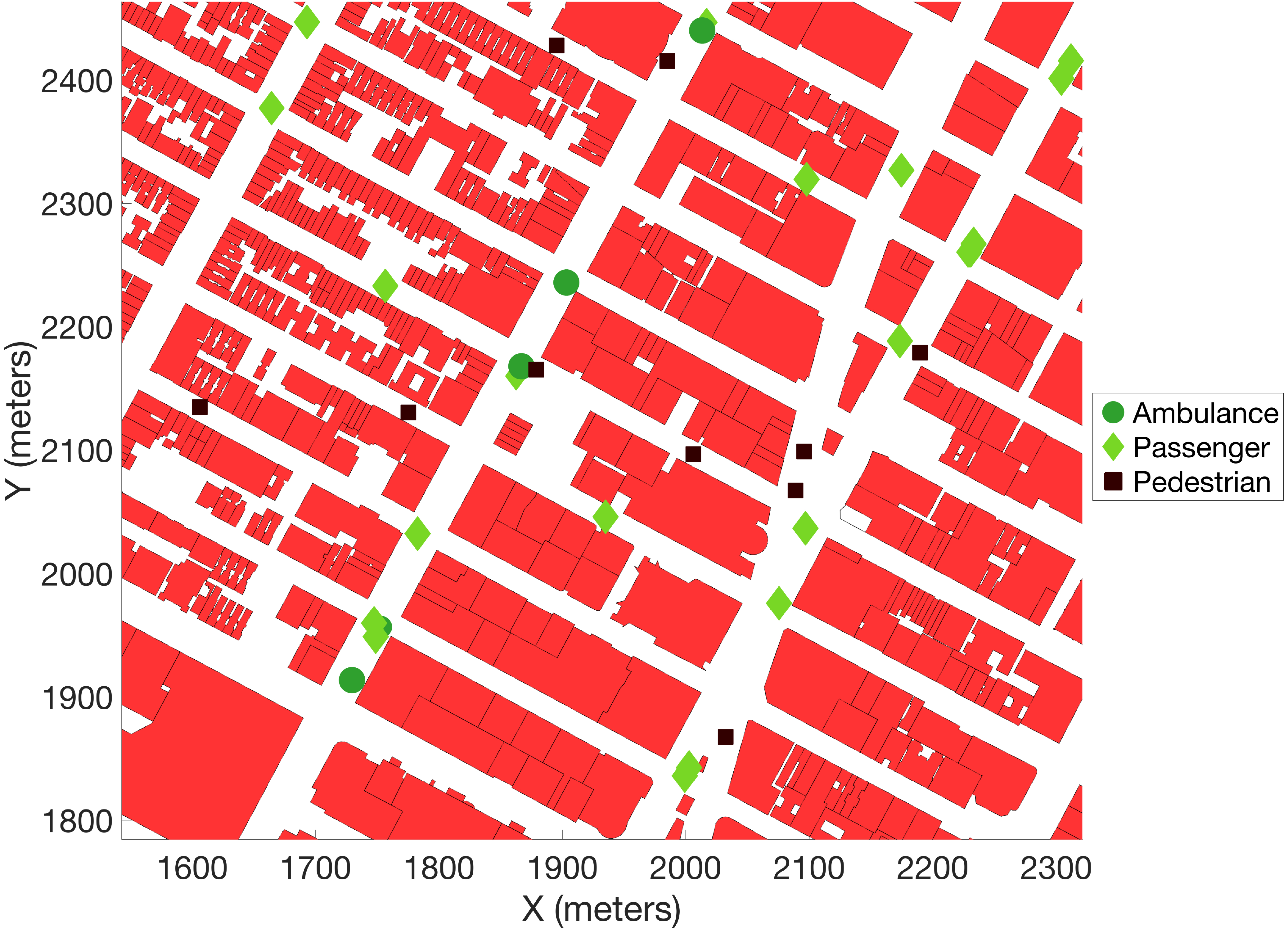}}
\end{minipage}
\begin{minipage}{.62\linewidth}
\centering
\subfloat[Example of the adaptive TX power for macrocell BSs and four different time instances.]{\label{fig:heatmapAll}\includegraphics[width=0.98\linewidth]{{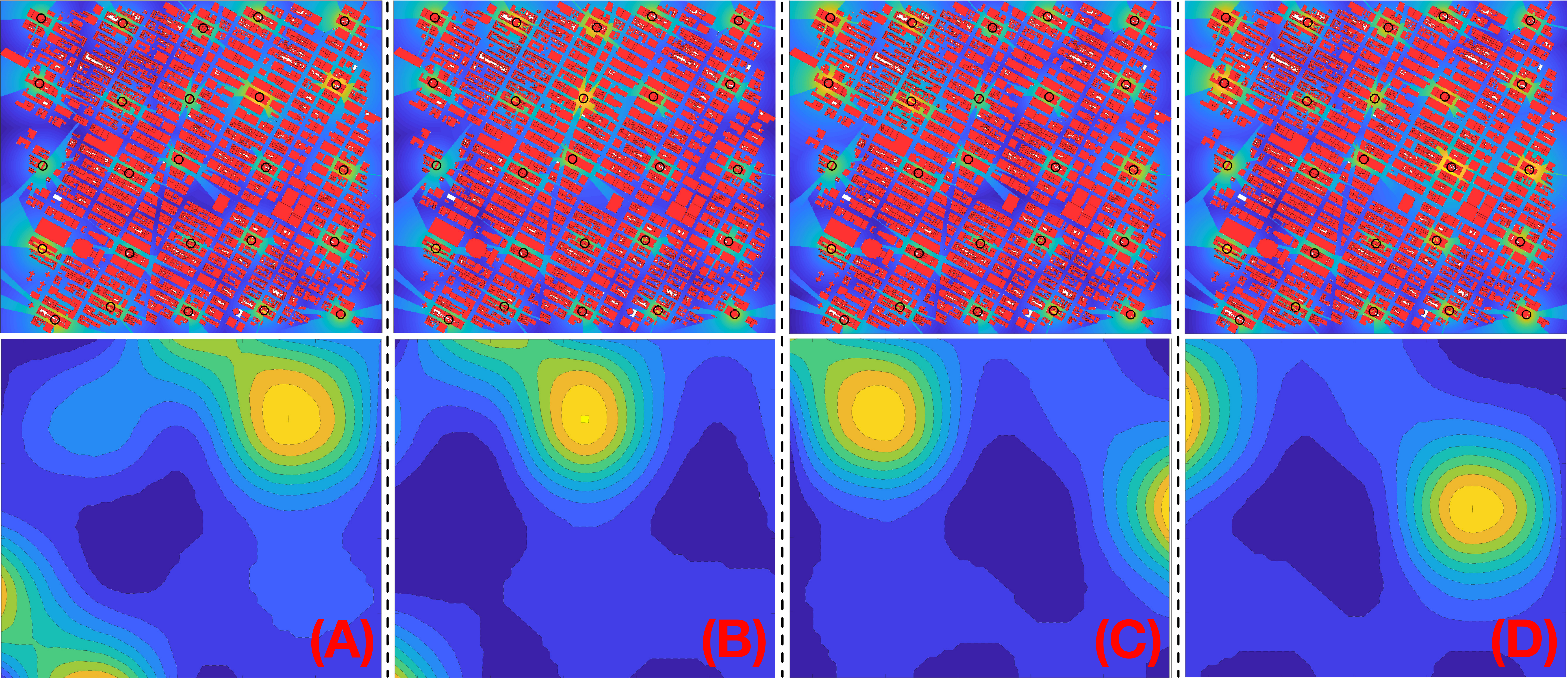}}}
\end{minipage}\par\medskip
\caption{An intelligent agent adapts the TX power based on the user density. Fig.~~\ref{fig:mobility} shows an example mobility trace of the two generated vehicle types and pedestrians. Fig.~\ref{fig:heatmapAll}-top shows the heatmap with the perceived RSSI and Fig.~\ref{fig:heatmapAll}-bottom shows the user densities. }
\label{fig:all}
\end{figure*}

For our use-case, we discretized the map into $\SI{200}{\meter} \times \SI{200}{\meter}$ areas (as in Fig.~\ref{fig:tessellation}). Later, we find the highest and lowest user densities for all areas, calculated as the number of vehicles, pedestrians, and indoor users within this area. A snapshot of the different vehicles and pedestrians positions is shown in Fig.~\ref{fig:mobility}. Having these densities, we later adapt the TX power of our BSs to increase the perceived datarate. More specifically, we assume that the macrocell BSs in close proximity to the area with the lowest density is configured with a TX power of \SI{20}{\dBm}. On the other hand,  the BS in areas with the highest density is tuned at \SI{43}{\dBm} of TX power. Proportionally, we configure the rest of the BSs. Similarly, we configure the femtocell BSs as well using a TX power range between \SIrange{15}{25}{\dBm}. For the ground truth, we set all the BSs to the median value, i.e., \SI{31.5}{\dBm} for the macrocell BSs and \SI{20}{\dBm} for the femtocell ones.

\begin{figure}[t]     
\centering
    \includegraphics[width=1\columnwidth]{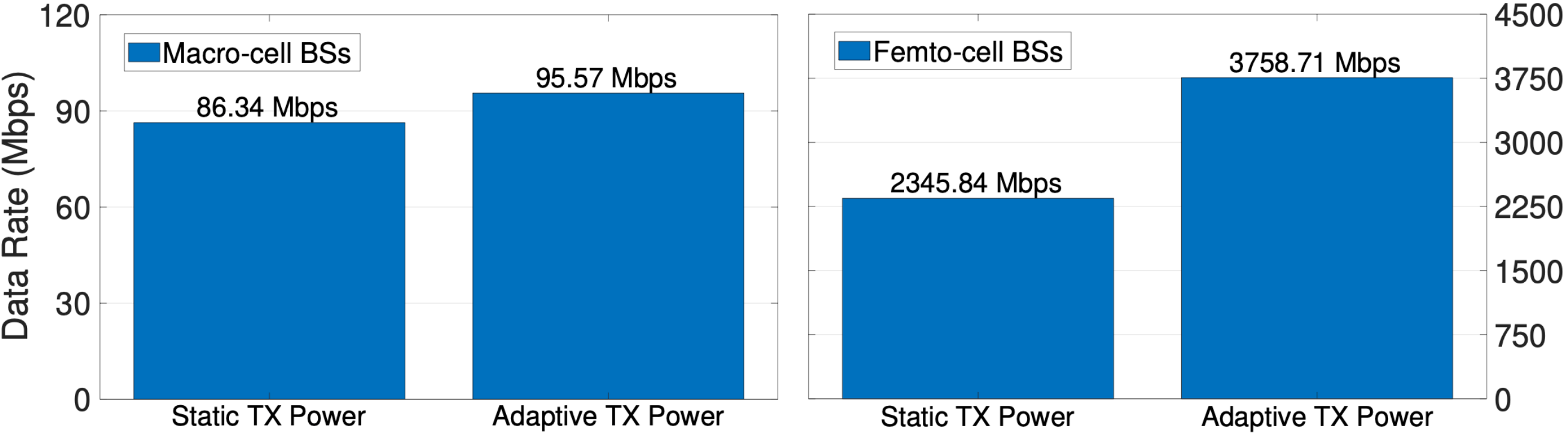}
    \caption{Comparison between the static and adaptive TX power scenarios.}
    \label{fig:datarate}
\end{figure}

In Fig.~\ref{fig:all}, we see four instances of the executed scenario. For each one, the user density was calculated and visualized for the entire city (Fig.~\ref{fig:heatmapAll}-bottom). As shown, DRIVE provides progressively calculates changes the user density on the map. Fig.~\ref{fig:heatmapAll}-top shows the corresponding RSSI heatmaps for the macrocell plane. As discussed before, the TX power is adapted based on the density. In Fig.~\ref{fig:datarate}, we see the datarate for the static and the adaptive TX power scenarios. As expected, when the TX power is increased in areas with more users, the datarate is being increased as well. Even though this is a very simplistic use-case, it demonstrates very well the capabilities of DRIVE. As discussed before, our framework was designed to provide a flexible and fast way of interacting with intelligent agents, deploying policies, and evaluating performance improvements. Utilizing more sophisticated agents in the future, several C-ITS use-cases can be evaluated in an SoS-fashion.

\section{Conclusions and Future Work}\label{sec:conclusions}
\vspace{-1mm}
In this paper, we described DRIVE. DRIVE is a city-scale Digital Network Oracle for C-ITS experimentation. Our solution is developed with an SoS ecosystem in mind and provides a robust environment for large-scale C-ITS-related scenario evaluation. Allowing an easy and adaptive bidirectional interaction with real-world mobility traces and intelligent agents, DRIVE can be used for several use-cases spanning from cybersecurity within C-ITSs, to behavioral system analysis based on various conditions. Furthermore, being highly flexible, modular, and scalable can simulate city-scale realistic scenarios within minutes. Overall, DRIVE was designed to tackle the problems of the existing simulation frameworks and enhance the prototyping of intelligent solutions for pressing concerns of the next-generation C-ITSs. In the future, DRIVE will be integrated with ML intelligent agents, and several C-ITS use-cases will be evaluated.

\vspace{-1mm}
\section*{Acknowledgment}
\vspace{-1mm}
This work is funded in part by Toshiba Europe Ltd., in part by the Next-Generation Converged Digital Infrastructures (NG-CDI) project, supported by BT Group and EPSRC (EP/R004935/1), and in part by the CAVShield project (Innovate UK, no. 133898).

\bibliographystyle{IEEEtran}
\bibliography{bib.bib,IEEEabrv}

\end{document}